\documentclass[twoside,twocolumn,9pt]{article}
\usepackage[utf8]{inputenc}
\usepackage{extsizes}
\usepackage[super,sort&compress,comma]{natbib} 
\usepackage[version=3]{mhchem}
\usepackage[left=1.5cm, right=1.5cm, top=1.785cm, bottom=2.0cm]{geometry}
\usepackage{balance}
\usepackage{times,mathptmx}
\usepackage{sectsty}
\usepackage{graphicx} 
\usepackage{lastpage}
\usepackage[format=plain,justification=justified,singlelinecheck=false,font={stretch=1.125,small,sf},labelfont=bf,labelsep=space]{caption}
\usepackage{float}
\usepackage{fancyhdr}
\usepackage{fnpos}
\usepackage[english]{babel}
\addto{\captionsenglish}{%
  
}
\usepackage{amsmath}
\usepackage{amssymb}

\usepackage{array}
\usepackage{droidsans}
\usepackage{charter}
\usepackage[T1]{fontenc}
\usepackage[usenames,dvipsnames]{xcolor}
\usepackage{setspace}
\usepackage[compact]{titlesec}
\usepackage{hyperref}
\usepackage{epstopdf}

\definecolor{cream}{RGB}{222,217,201}

\begin{document}

\pagestyle{fancy}
\thispagestyle{plain}
\fancypagestyle{plain}{

\renewcommand{\headrulewidth}{0pt}
}

\makeFNbottom
\makeatletter
\renewcommand\LARGE{\@setfontsize\LARGE{15pt}{17}}
\renewcommand\Large{\@setfontsize\Large{12pt}{14}}
\renewcommand\large{\@setfontsize\large{10pt}{12}}
\renewcommand\footnotesize{\@setfontsize\footnotesize{7pt}{10}}
\makeatother

\renewcommand{\thefootnote}{\fnsymbol{footnote}}
\renewcommand\footnoterule{\vspace*{1pt}%
\color{cream}\hrule width 3.5in height 0.4pt \color{black}\vspace*{5pt}} 
\setcounter{secnumdepth}{5}

\makeatletter 
\renewcommand\@biblabel[1]{#1}            
\renewcommand\@makefntext[1]%
{\noindent\makebox[0pt][r]{\@thefnmark\,}#1}
\makeatother 
\renewcommand{\figurename}{\small{Fig.}~}
\sectionfont{\sffamily\Large}
\subsectionfont{\normalsize}
\subsubsectionfont{\bf}
\setstretch{1.125} 
\setlength{\skip\footins}{0.8cm}
\setlength{\footnotesep}{0.25cm}
\setlength{\jot}{10pt}
\titlespacing*{\section}{0pt}{4pt}{4pt}
\titlespacing*{\subsection}{0pt}{15pt}{1pt}

\fancyfoot{}
\fancyfoot[RO]{\footnotesize{\sffamily{1--\pageref{LastPage} ~\textbar  \hspace{2pt}\thepage}}}
\fancyfoot[LE]{\footnotesize{\sffamily{\thepage~\textbar\hspace{3.45cm} 1--\pageref{LastPage}}}}
\fancyhead{}
\renewcommand{\headrulewidth}{0pt} 
\renewcommand{\headrulewidth}{0pt}
\setlength{\arrayrulewidth}{1pt}
\setlength{\columnsep}{6.5mm}
\setlength\bibsep{1pt}

\makeatletter 
\newlength{\figrulesep} 
\setlength{\figrulesep}{0.5\textfloatsep} 

\newcommand{\topfigrule}{\vspace*{-1pt}%
\noindent{\color{cream}\rule[-\figrulesep]{\columnwidth}{1.5pt}} }

\newcommand{\botfigrule}{\vspace*{-2pt}%
\noindent{\color{cream}\rule[\figrulesep]{\columnwidth}{1.5pt}} }

\newcommand{\dblfigrule}{\vspace*{-1pt}%
\noindent{\color{cream}\rule[-\figrulesep]{\textwidth}{1.5pt}} }

\makeatother


\twocolumn[
  \begin{@twocolumnfalse}
\vspace{3cm}
\sffamily
\begin{tabular}{m{4.5cm} p{13.5cm} }

 &\noindent\LARGE{\textbf{Effect of particle deformability on shear thinning in a 3D channel}} \\
\vspace{0.3cm} & \vspace{0.3cm} \\

 & \noindent\large{Danilo P. F. Silva,$^{\ast}$\textit{$^{a,b}$} Rodrigo C. V. Coelho,\textit{$^{a,b}$} Margarida M. Telo da Gama,\textit{$^{a,b}$} and Nuno A. M. Araújo \textit{$^{a,b}$}} \\\\

&\noindent\normalsize{
Soft particles suspended in fluids flowing through microchannels are often encountered in biological flows such as cells in vessels.  They can deform under flow or when subject to mechanical stresses as they interact with themselves. Deformability adds distinct characteristics to the nature of the flow of these particles. We simulate deformable particles suspended in a fluid at a high volume fraction flowing through a 3D wetting channel. We find a discontinuous shear thinning transition, which depends on the particle deformability. The capillary number is the main parameter that controls the transition. We also notice significant differences in the velocity profiles when compared to their 2D counterpart. To obtain these results, we improved and extended to 3D a multicomponent Lattice Boltzmann method which prevents the coalescence between the droplets.
} 

\\

\end{tabular}

 \end{@twocolumnfalse} \vspace{0.6cm}

  ]

\renewcommand*\rmdefault{bch}\normalfont\upshape
\rmfamily
\section*{}
\vspace{-1cm}


\footnotetext{\textit{$^{a}$~Centro de Física Teórica e Computacional, Faculdade de Ciências, Universidade de Lisboa, 1749-016 Lisboa, Portugal; E-mail: dpsilva@fc.ul.pt}}
\footnotetext{\textit{$^{b}$~Departamento de Física, Faculdade de Ciências,
Universidade de Lisboa, P-1749-016 Lisboa, Portugal.}}

\footnotetext{\dag~Electronic Supplementary Information (ESI) available: Description of the supplementary movies and supplementary figures. See DOI: 00.0000/00000000.}




\section{Introduction}

Particles suspended in fluids flowing through channels are often encountered in industrial and natural processes. The flow of blood, microfluidic devices and drug delivery are typical applications~\cite{Xu_2009}. For instance,  microfluidics has matured and provided new techniques for the control of soft suspensions, particularly in cell manipulation and cell studies~\cite{li_2014}. This has allowed researchers to simulate microgeometries with enhanced control of cells in a variety of flows such as cell adhesion, cell suspensions and cell sorting~\cite{young_fundamentals_2010,mehling_microfluidic_2014}.

Generally, these complex suspensions consist of soft deformable particles which exhibit different flow behavior when compared to their hard counterparts~\cite{koumakis2012,vaart2013}. Systems where particles deform due to hydrodynamic interactions have not been systematically studied and a deeper understanding of the key physical properties that control such systems is desirable. 

There has been a growing effort to model the collective behavior of soft suspensions in channels and the role of particle deformability in such systems. Several numerical methods have been used to study soft particles in microchannels and other confined geometries. A well known method is the boundary integral method (BIM) and its variants, specially used in the study of multiphase problems~\cite{Hou_2001}. A major challenge to the use of these methods is how to accurately and effectively include surface tension. However, BIM is applicable to certain types of flows such as Stokes or potential flows. A popular variant of BIM is the method of interfacial dynamics~\cite{POZRIKIDIS_2001}, which is suitable for the simulation of incompressible soft particles. Using this method, it has been shown that ordered droplets driven by a pressure difference in a periodic channel accumulate at different regions of the channel depending on the viscosity ratio between the droplet and the surrounding fluid~\cite{Zhou1994}. When the viscosity ratio is equal to unity they accumulate at the center of the channel but, for higher values of the viscosity ratio the droplets may accumulate at the walls leaving a droplet free zone in the middle. A more detailed study on particle migration and concentration was conducted using Stokesian Dynamics in 3D~\cite{nott_brady_1994}, which allows access to much smaller Reynolds numbers ($\rm Re \ll 1$). However, it appears to be restricted in terms of applications since it is was specifically developed for spherical particles in Stokes flow. If one wants to study deformable non-spherical particles other methods such as the Lattice Boltzmann method (LBM) provide a way to include surface tension. LBM is commonly used to study deformable particles under different conditions~\cite{Dunweg_2009,kruger_2011,chen_2014}. For instance, it has been used to investigate how 3D particle migration and distribution is affected by deformability and inertia at different Reynolds numbers, in particular how distinct flow focusing emerges at increasing Reynolds number for strongly deformable particles~\cite{kruger_kaoui_harting_2014}. In those limits, a non-monotonic behavior of the apparent viscosity of the suspension was reported.

The flow of suspensions exhibits a wide variety of non-Newtonian behavior such as shear thinning. Shear thinning has been observed in experimental studies of hard suspensions such as glass balls in oil~\cite{brown_2009} and cornstarch particles in water~\cite{fall_2015}. A striking phenomenon is the observation of shear thinning in noncolloidal suspensions reported in numerous experiments~\cite{isidro,shao_2013,lin_2014,singh_nott_2003}. In these experiments, particles larger than 20 $\mu$m were used and the effects of particle migration, sedimentation and confinement were investigated where Brownian effects are negligible.

Previous experimental studies have shown that shear thinning in soft suspensions depends strongly on the size of the particles~\cite{otsubo_effect_1994,Pal1996,Pal2000}. These experiments typically involve oil droplets in water studied over a large range of volume fractions. In parallel with experiments, shear thinning has also been reported in emulsions through numerical simulations for example, using interface tracking methods~\cite{loewenberg_hinch_1996}. The collective dynamics of a large number of deformable particles in microchannels was also investigated~\cite{Doddi2009} via 3D simulations using an immersed boundary method. The trajectories and the plug-flow profile of the particles were analyzed as a function of the deformability and channel size. Other works ~\cite{lazaro2014_I,lazaro2014_II} investigated the effect of elasticity and confinement of red blood cell (RBC) suspensions for different cell rigidity. They observed shear thinning behavior and reported that at high enough capillary numbers the effective viscosity of the suspension converges to the solvent viscosity. These studies suggest that a key parameter when investigating the flow of deformable particles is the capillary number.

Recently, a discontinuous shear thinning behavior was reported in the flow of deformable droplets in 2D ~\cite{Foglino2017}. This behavior is associated with a nonequilibrium transition between a "soft" and a "hard" phase, that depends on the area fraction of droplets in suspension and the applied pressure difference. Strong shear thinning behavior was reported at higher area fraction and it was further revealed that when the area fraction is higher than 0.5, the shear thinning becomes discontinuous i.e. there is a jump in the viscosity at a critical value of the forcing. In addition, the velocity (along the flow direction) was measured and it was reported that the velocity of the droplets remains close to that of the surrounding fluid throughout the channel. Simulations were performed using a 2D hybrid LBM. A more recent study ~\cite{fei_2020} investigated and confirmed the robustness of this discontinuous behavior. This study reported additional discontinuous jumps in viscosity over a larger parameter range, showing how the discontinuous shear thinning is preserved at lower values of the confinement, defined as the ratio between channel height and droplet radius. To do so, they simulated a large 2D system with up to 500 droplets ($\approx$ 0.85 of area fraction). 

While this discontinuous shear thinning behavior has been investigated over a wide parameter range in 2D, these effects have not yet been identified and tested in 3D geometries. In particular, how deformability affects the discontinuous shear thinning and the overall fluid flow in 3D systems. This is relevant because real fluid flow is 3D and parameters such as the velocity and pressure vary in the three coordinate directions. An important example of this is the flow of RBC which may deform due to geometrical constraints in vessels and display effects only observed in 3D environments~\cite{fedosov_2014}. In this paper, we simulate particles in 3D flows and investigate the discontinuous shear thinning behavior as a function of the surface tension, which is related to the particle deformability. Here, we improve and extend to 3D a multicomponent LBM with frustrated coalescence~\cite{benzi_mesoscopic_2009}. We analyze the shear thinning for different flow conditions and particle deformability. We also discuss the velocity profiles and compare them to their 2D counterparts.

The paper is organized as follows. The LBM is developed in Sec. 2 and validated in Sec. 3. In Sec. 4, the effect of particle deformability is studied for a dense suspension flow in a 3D cylindrical channel. We highlight differences and similarities between the results in 2D and 3D flows. In Sec. 5, we make some final observations.

\section{Method}
\label{method-sec}

In what follows, we develop a 3D multicomponent model with frustrated coalescence. In the 3D model for flexible particles, the motion of the fluid is represented by a set of distribution functions $f_{k, i}(\mathbf{x}, t)$  at position $\mathbf{x}$ and time $t$, where the subscripts $k$ and $i$ denote the fluid component and discrete velocity directions, respectively. The time evolution of $f_{k, i}\left(\mathbf{x},t\right)$ is given by the discrete Boltzmann equation,

\begin{equation}
f_{k, i}\left(\mathbf{x}+\mathbf{c}_{i}, t+1\right)-f_{k, i}(\mathbf{x}, t)=-\frac{1}{\tau_{k}}\left[f_{k, i}(\mathbf{x}, t)-f_{k, i}^{e q}(\mathbf{x}, t)\right]+\mathcal{F}_{k, i} ,
\label{eq:lb_eq}
\end{equation}
where $\tau_{k}$ is the relaxation time for each component and $\mathcal{F}_{k, i}$ is the forcing term. Unless otherwise stated, we express our results in lattice units (l.u.), which means that $\Delta x$ and $\Delta t$ are the length and time units. The equilibrium distribution is given by

\begin{equation}
f_{k, i}^{e q}=\rho_{k} w_{i}\left[1+\frac{\mathbf{u}^{e q} \cdot \mathbf{c}_{i}}{c_{s}^{2}}+\frac{\left(\mathbf{u}^{e q} \cdot \mathbf{c}_{i}\right)^{2}}{2 c_{s}^{4}}-\frac{\left(\mathbf{u}^{e q}\right)^{2}}{2 c_{s}^{2}}\right] ,
\label{eq:equillibirum_dist}
\end{equation}
where $c_s$ is the speed of
sound. The quadrature used in the streaming step is the D3Q41 lattice (see Table \ref{d3q41-table} in Appendix for the vectors $\mathbf{c}_{i}$ and weights $w_{i}$). $\mathbf{u}^{e q}$ is the equilibrium velocity given by

\begin{equation}
\mathbf{u}^{e q}=\frac{\sum_{k} \frac{\rho_{k} \mathbf{u}_{k}}{\tau_{k}}}{\sum_{k} \frac{\rho_{k}}{\tau_{k}}} ,
\end{equation}
where $\rho_{k} \mathbf{u}_{k}$ is the $k$th component of momentum. The Guo forcing scheme~\cite{Guo_2002} was adopted to implement the forces acting in a fluid because it leads to a viscosity-independent surface
tension

\begin{equation}
\mathcal{F}_{k, i}=\left(1-\frac{1}{2 \tau_{k}}\right) w_{i}\left[\frac{\mathbf{c}_{i}-\mathbf{u}^{e q}}{c_{s}^{2}}+\frac{\left(\mathbf{c}_{i} \cdot \mathbf{u}^{e q}\right) \mathbf{c}_{i}}{c_{s}^{4}}\right] \cdot \mathbf{F}_{k},
\label{eq:guo_force}
\end{equation}
where $\mathbf{F}_{k}$ is the sum of all the forces. The equilibrium velocity is the same as in Eq.\eqref{eq:equillibirum_dist} \cite{guo_zheng_shi_2002}. Also,

\begin{equation}
\rho_{k}=\sum_{i=0}^{40} f_{k, i}, \quad \rho_{k} \mathbf{u}_{k}=\sum_{i=0}^{40} f_{k, i} \mathbf{c}_{i} + \frac{ \mathbf{F}_{k} }{2}.
\end{equation}
Lastly, the barycentric velocity $\mathbf{u}$ of the fluid mixture i.e. the physical velocity is given by
\begin{equation}
\mathbf{u}=\frac{\sum_{k} \rho_{k} \mathbf{u}_{k}}{\rho}, \quad \rho=\sum_{k} \rho_{k} ,
\end{equation}
where $\rho$ is the total density of the fluid mixture.

There are three internal forces acting on the fluid: an attractive short-range one, a repulsive midrange force, and a repulsive force between the components. The midrange force is what prevents droplet coalescence and, in that case,  we use the D3Q39 lattice (see Table \ref{d3q39-table} in appendix). For the attractive short-range and repulsive forces between the components, we use the D3Q41 lattice. The competing forces acting on the same component are

\begin{equation}
\begin{split}
    \mathbf { F } _ { k } ^ { c } = - G _ { k , 1 } \psi _ { k } ( \mathbf { x } ) \sum _ { i = 0 } ^ { 40 } w _ { i } \psi _ { k } \left( \mathbf { x } + \mathbf { c } _ { i } \right) \mathbf { c } _ { i }\\ - G _ { k , 2 } \psi _ { k } ( \mathbf { x } ) \sum _ { j = 0 } ^ { 38 } p _ { j } \psi _ { k } \left( \mathbf { x } + \mathbf { c } _ { j } \right) \mathbf { c } _ { j },
    \end{split}
    \label{eq:competing_force}
\end{equation}
where the pseudo-potential is $\psi _ { k } \left( \rho _ { k } \right) = \rho _ { 0 } \left( 1 - e ^ { - \rho _ { k } / \rho _ { 0 } } \right)$ with a uniform reference density $\rho _ { 0 }$. Additionally, $G _ { k , 1 }$ and $G _ {k , 2 }$ are the self-interaction forces within each component. The repulsive forces between the components are implemented as usual~\cite{kruger2016}

\begin{equation}
    \mathbf { F } _ { k } ^ { r } = - \rho _ { k } ( \mathbf { x } ) \sum _ { \overline { k } } G _ { \overline { k } k } \sum _ { i = 0 } ^ { 40 } w _ { i } \rho _ { \overline { k } } \left( \mathbf { x } + \mathbf { c } _ { i } \right) \mathbf { c } _ { i }
    \label{repulsive_force}.
\end{equation}
The interaction force with the solid boundary is given by

\begin{equation}
\mathbf { F } _ { k } ^ { s } = - G _ { k s} \rho _ { k } ( \mathbf { x } ) \sum _ { i = 0 } ^ {solid} w _ { i } s \left( \mathbf { x } + \mathbf { c } _ { i }\right) \mathbf { c } _ { i },
\label{solid_force}
\end{equation}
where $s(\mathbf { x })$ is the switch function, which takes the values 0 and 1 for fluid and solid nodes, respectively, and $G_{ k s}$ is the interaction strength between fluid component $k$ and the solid boundary. The subscript $k$ represents the component A, B, etc.

The total force $\mathbf{F}_{k}$ in Eq.\eqref{eq:guo_force} is the sum of the external body forces $\mathbf{F}_{k}^{b}$ (e.g., gravity, which we neglect here) the internal forces and the solid boundary interaction force: $\mathbf{F}_{k}=\mathbf{F}_{k}^{b}+\mathbf{F}_{k}^{c}+\mathbf{F}_{k}^{r}+\mathbf{F}_{k}^{s}$. For simplicity, we assume: $G_{A, 1}=G_{B, 1}$, $G_{A, 2}=G_{B, 2}$, $G_{A, B}=G_{B, A}$. The parameters $G_{A, 1}$, $G_{B, 2}$, etc are the strength coefficients of the interaction forces. More specifically, a positive (or negative) strength coefficient represents a repulsive (or attractive) interaction. To have the desired repulsive and attractive forces: $G_{k, 1}<0$, $G_{k, 2}>0$, and $G_{k, \overline{k}}>0$. In the competing force $\mathbf{F}_{k}^{c}$, the attractive force must overcome the repulsive force to form droplets so, we set $\left|G_{k, 1}\right|>\left|G_{k, 2}\right|$.

\begin{figure}[H]
\center
\includegraphics[width=1.00\linewidth]{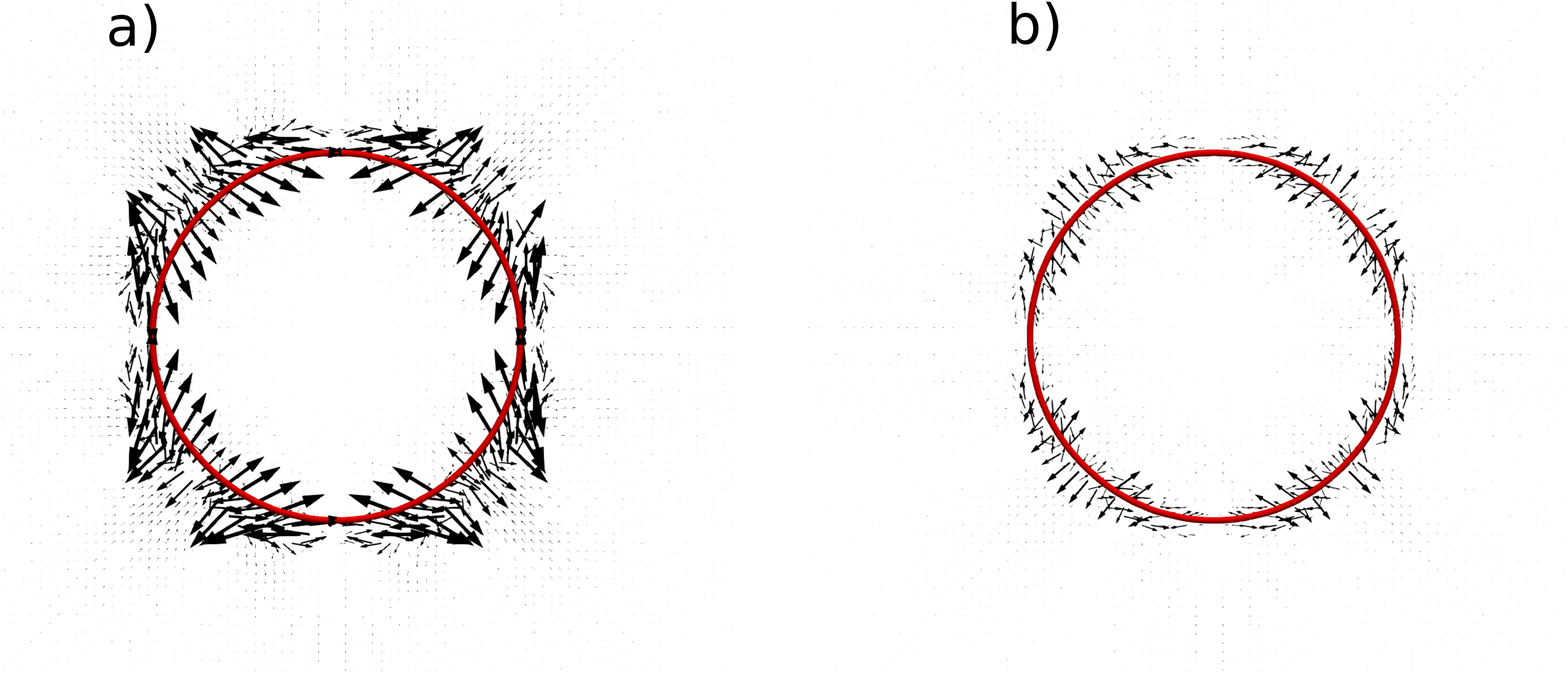}
\caption{Spurious velocities (arrows) around a circular interface (red) of radius $r = 20$ at equilibrium for $G_{k, 1} = -7.9$, $G_{k, 2} = 4.9$. The largest spurious velocity for (a) D3Q19 is 0.017 l.u. and for the (b) D3Q41 lattice it is 0.0092 l.u.. The velocity field is on the same scale in both images.}
\label{spurious-fig}
\end{figure}

Multiphase and multicomponent LBM suffer from spurious velocities which are caused by an imbalance between the interaction forces. These velocities can increase if the viscosity ratio deviates much from one or if the surface tension is high. Thus, the simulations might become unstable for realistic physical parameters. In 2D, the lattice used for the midrange repulsive force was the D2Q25~\cite{falcucci_lattice_2007,dollet_two-dimensional_2015}. In 3D the equivalent of the D2Q25 lattice is the D3Q125 lattice, however this lattice does not satisfy all the isotropic requirements and exhibits high spurious velocities at the interface which leads to droplet coalescence. The D3Q39  was used for the midrange force instead. A non-physical effect caused droplets to become "glued" along the direction of the diagonal vectors of the lattice. We also observed this effect in the usual 2D models~\cite{falcucci_2007,benzi_2007,sbragalia_2007}. Peng \textit{et al.}~\cite{peng_isotropy_2019} showed that using higher isotropic lattices in the streaming process reduces significantly the spurious velocities. In our simulations, we use an 8th order isotropic lattice in the streaming step. By contrast, the D3Q19 has only 4th order isotropy. In Fig.~\ref{spurious-fig}, we compare the spurious velocities when these two lattices are used in the streaming step.  We note that it reduces spurious velocities by nearly one half. Notice in Fig.~\ref{spurious-fig} that spurious velocities are stronger along the diagonal direction of the grid, which explains why the droplets stick to each other at these points but not along the grid axes.

\section{Numerical Tests}
\label{model-sec}

In this section, we analyze the properties of the 3D multicomponent method. We measure the surface tension, which is related to the deformability of the soft particles and we analyze the disjoining pressure.

\begin{figure}[H]
\center
\includegraphics[width=1.00\linewidth]{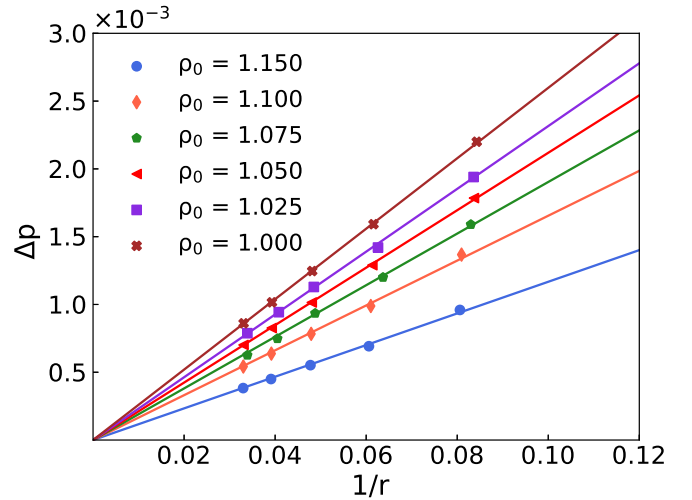}
\caption{Pressure difference $\Delta p$ as a function of the inverse of droplet radius $r$. We simulate a static droplet for a fixed $\rho_0$ (reference density) and let it reach the steady state. The dots represent the pressure difference inside and outside the droplet in our simulations. The solid lines are linear fits corresponding to Eq. \ref{eq:surface_tension}, where the slope is $\gamma$.}
\label{laplace3d-fig}
\end{figure}

To measure the surface tension, we perform the Laplace test. At equilibrium, the curved interface of a static droplet increases the pressure inside the droplet. The radius, pressure difference and surface tension must satisfy the Young-Laplace equation \cite{kruger_lattice_2017}:

\begin{equation}
\Delta p=p_{\text{in}} - p_{\text{out}} =\frac{\gamma}{r} ,
\label{eq:surface_tension}
\end{equation}
where $r$ is the radius of the droplet, $p_{\rm in}$ and $p_{\rm out}$ are the pressures inside and outside the droplet, respectively. $\gamma$ is the surface tension. To test Laplace’s law, a series of LBM simulations with different values for the droplet radius were performed. As seen in Fig.~\ref{laplace3d-fig}, the pressure difference $\Delta p$ increases linearly with the inverse of the droplet radius $r$ (curvature). To vary the surface tension, we fix the parameters $G_{k, 1}$ and $G_{k, 2}$ and change the uniform reference density $\rho_{0}$ (effectively varying the interaction forces). The solid lines represent results calculated from Laplace’s law, and the dots are obtained from the simulations. The slope of the solid lines is $\gamma$. The simulation results are consistent with Laplace’s law. The values of $\gamma$ are shown in Table~\ref{gamma-table}.

\begin{figure}[H]
\center
\includegraphics[width=0.96\linewidth]{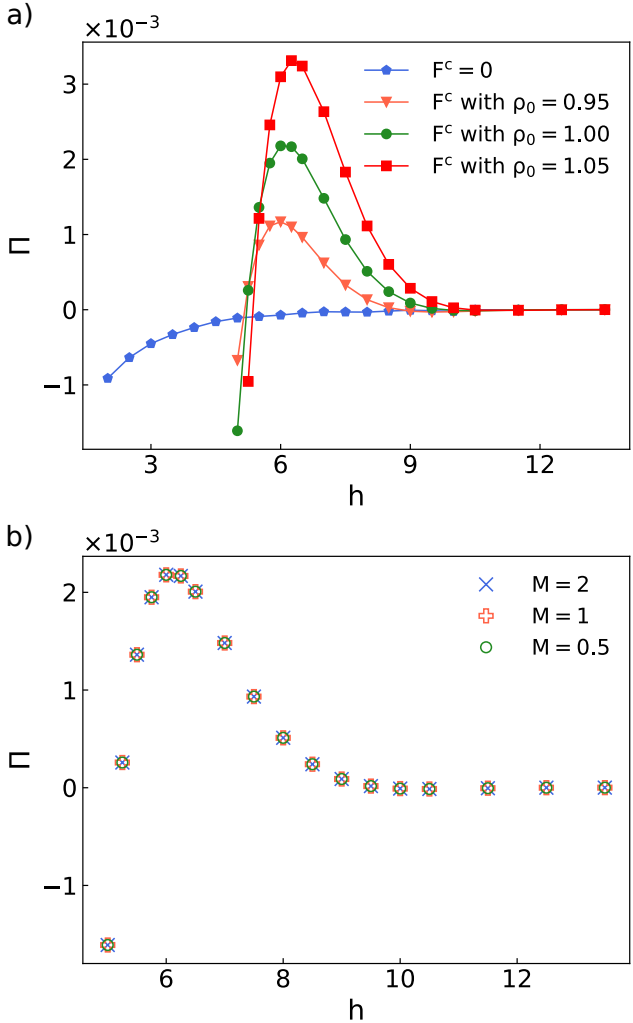}
\caption{Disjoining pressure $\Pi$ as a function of the distance h between two flat interfaces with fixed $G_{k,1} = -3.5$, $G_{k,2} = 2.5$ achieved with competing forces $\rm F^c$ given by Eq.\eqref{eq:competing_force} which provides the repulsive force to avoid coalescence. A positive value indicates repulsion while a negative value indicates attraction. (a) Disjoining pressure for different values of the reference density $\rho_0$. (b) Disjoining pressure for different values of the viscosity ratio M with $\rho_0 = 1.00$.}
\label{disjointpressure-fig}
\end{figure}

When the parameter $G_{AB}$ exceeds a certain threshold it gives rise to stable interfaces between fluids A and B with positive surface tension. However, when the droplets approach each other, a thin film is formed leading to coalescence. The phase separation promotes negative disjointing pressure and thus the competing interactions prevent coalescence of neighboring droplets and give rise to positive disjoining pressure as seen in Fig.~\ref{disjointpressure-fig}. This mechanism has been used in other studies to simulate non-coalescing droplets~\cite{benzi_mesoscopic_2009,sbragaglia_emergence_2012,benzi_internal_2015}. The disjoining pressure is also independent of the viscosity ratio $\rm M$ defined as the viscosity of the droplet over the viscosity of the surrounding fluid~\cite{fei_mesoscopic_2018}.

\begin{table}[h!]
\caption{Surface tension $\gamma$ for different reference densities $\rho_0$ in lattice units. The values of $\gamma$ are obtained from linear fits in Fig.~\ref{laplace3d-fig}.}
\begin{tabular} {|*{7}{l|}}
\hline
$\rho_0$ & 1.000 & 1.025 & 1.050 & 1.075 & 1.100 & 1.150 \\ \hline
$\gamma$ & 0.026 & 0.023 & 0.021 & 0.019 & 0.017 & 0.012\\ \hline
\end{tabular}
\label{gamma-table}
\end{table}

\section{Flow of deformable particles}
\label{results-sec}

\begin{figure}[H]
\center
\includegraphics[width=0.77\linewidth]{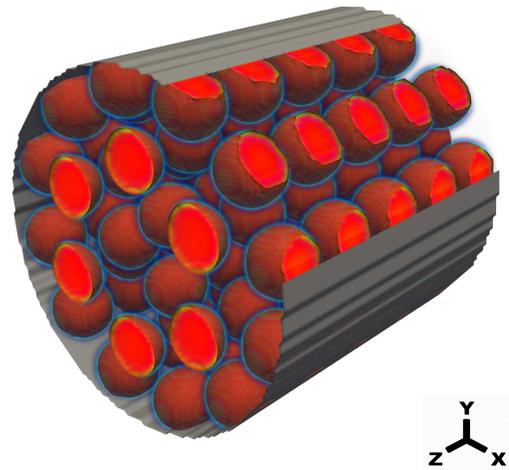}
\caption{Initial setup. 105 droplets non-close packed with volume fraction of 0.61. A force $\rm F$ is imposed in the $z$-direction and varied. The walls have a neutral wetting condition. The droplets (red) inside the cylindrical channel (grey). The system is periodic in the direction of the flow. The parameters for the interactions are $G_{k,1} = -7.9$, $G_{k,2} = 4.9$ and the viscosity of both droplet and surrounding fluid is set to one.}
\label{snapshots_initial-fig}
\end{figure}

To understand the effect of particle deformability on the overall flow of concentrated suspensions, we simulate a densely packed system of 105 droplets of radius $r = 8$ lattice units in a 3D cylindrical channel with periodic boundary conditions in the $z$-direction (see Fig.~\ref{snapshots_initial-fig}). The droplets are hexagonally non-close packed resulting in a total volume fraction of 0.61. The height and radius of the cylinder are 95 and 47.5 lattice units, respectively. We impose no-slip boundary conditions in velocity and a neutral wetting boundary condition which creates a layer of wetting droplets along the wall. The system is driven by a fixed external force. The viscosity of the surrounding fluid and of the droplet's fluid is set to one in lattice units. Figure~\ref{snapshots_initial-fig} shows a snapshot of the initial setup.

\begin{figure*}[thb]
\center
\includegraphics[width=1.0\linewidth]{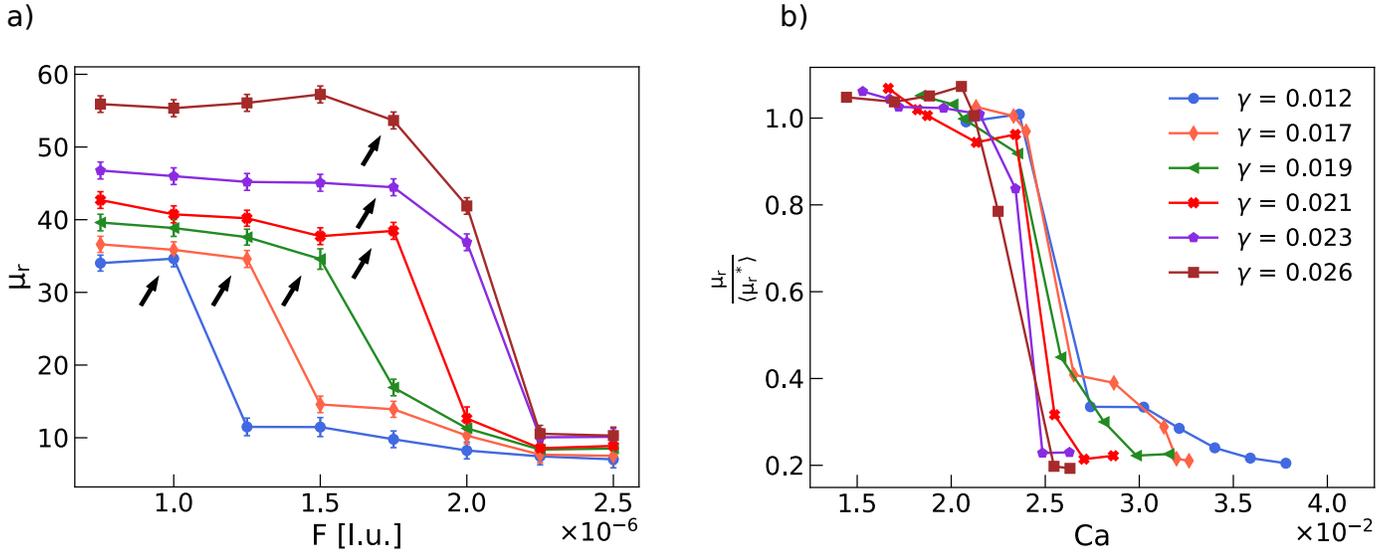}
\caption{Discontinuous shear thinning for different values of $\gamma$. (a) Discontinuous shear thinning occurs for progressively larger values of the external force which drives the flow. The black arrow indicates the threshold viscosity $\mu_{r}^{*}$ before the jump. (b) The relative viscosity $\mu_{r}$ (normalized by the threshold viscosity $\mu_{r}^{*}$) as a function of the capillary number $\rm Ca$. Simulations were carried out for different values of surface tension $\gamma$.}
\label{discontinuous-fig}
\end{figure*}

To study the effect of particle deformability we first vary the external force $\rm F$ along the z-direction and measure the effective relative viscosity $\mu_r = \mu/\mu_0$, where $\mu$ is the (apparent) dynamic viscosity of the fluid with droplets and $\mu_0$ is the dynamic viscosity of the fluid without droplets. The effective viscosity in terms of the volume flow rate
\begin{equation}
\mu_r = \frac{Q_0}{Q},
\label{effectiverelativeviscosity}
\end{equation}
is measured where $Q_0$ is the flow rate of the fluid without droplets and $Q$ is the flow rate with droplets. The flow rate $Q$ is defined as $Q =  \int_A v_z \,dA$ where $v_z$ is the axial velocity and A is the cross-sectional area. A time average of $Q$ is taken. This measurement of $\mu_r$ was obtained for different values of surface tension $\gamma$ (by varying $\rho_0$). The surface tension may be used as a measure of the particle deformability, however, since we compare forces resulting from fluid motion with forces resulting from surface tension, the capillary number $\rm Ca$ $= \mu_0 v/\gamma$  is a more appropriate dimensionless parameter, where $v$ is the characteristic velocity (taken as the maximum velocity of the fluid).

We study the relation between the effective viscosity and the surface tension. We plot the dependence of $\mu_r$ on the surface tension in Fig.~\ref{discontinuous-fig}. In Fig.~\ref{discontinuous-fig}a we see the discontinuous behavior which happens at larger values of $\rm F$ as we increase the surface tension. In addition, we see that the viscosity curves collapse when plotted against the capillary number. The discontinuous drop of the viscosity suggests shear thinning. For soft particles deformability has been shown to promote shear thinning analogous to cells flowing in microenvironments~\cite{lanotte_red_2016}.  Shear thinning is also observed in non-Brownian hard-sphere suspensions~\cite{kroupa_slip_2017} and the decrease in viscosity has been studied under more controlled conditions for these types of particles~\cite{isidro,Cwalina_2014}. These conditions typically involve glass or polystyrene spheres, which have the advantage of better defined particle properties and well studied packings, suspended in a specific fluid mixture at fixed composition (e.g. negligible evaporation or moisture absorption).

\begin{figure}[H]
\center
\includegraphics[width=1.00\linewidth]{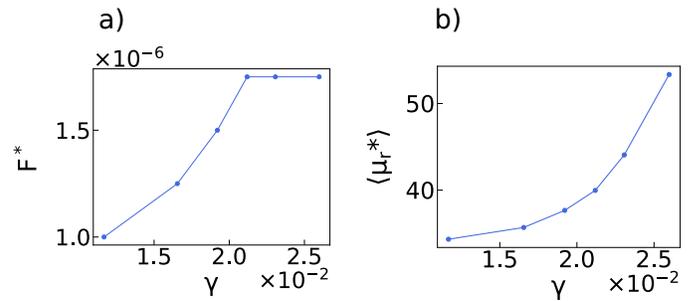}
\caption{(a) Threshold force $\rm F^*$ required to trigger discontinuous shear thinning as a function of $\gamma$. $\rm F^*$ increases roughly linearly with the droplets surface tension. (b) Corresponding threshold viscosity $\mu_{r}^{*}$ as function of $\gamma$.}
\label{fstarmustar-fig}
\end{figure}

We point out that the small bumps in some of the curves are the result of averaging when calculating the flow rate $Q$. Furthermore, we observe that at higher values of $\gamma$ the force required to trigger discontinuous shear thinning $ \rm F^*$ increases in a roughly linear fashion until it saturates. This is shown in Fig.~\ref{fstarmustar-fig}a. In fact, as we increase $\gamma$ the droplets become less flexible and eventually reach a state similar to that of hard particles. We expect in that limit shear thickening, as reported for example in Ref.~\cite{brown_2009} for hard suspensions (cornstarch in water and glass spheres in oil) between parallel plates (rotating top plate). Fig.~\ref{fstarmustar-fig}b shows that as the droplets become less deformable (increasing $\gamma$), the threshold viscosity $\mu_{r}^{*}$ increases.

The simulations were performed for a fixed value of the ratio between the channel diameter and the droplet radius. In 2D~\cite{fei_2020}, it is reported that the effective relative viscosity $\mu_r$ increases with this ratio. We did not vary this ratio but, we do not expect that the effect of confinement is qualitatively different in 3D.

The fully developed velocity profile across a cylindrical channel with no droplets can be computed analytically from the equations of fluid motion, the Navier-Stokes equations. For a cylindrical channel with circular cross-section of radius $R$, the solution is a parabolic velocity distribution~\cite{Schlichting(Deceased)2017}. The velocity distribution for droplets suspended in flow is, however, more complex. The flow in this channel is described by small characteristic dimensions and velocities. Consequently, the flow is characterized by low Reynolds number $\rm Re =$ $\frac{\rho_0 v D_{x}}{\mu_o}$, where $D_x$ is the diameter of the cylinder and is laminar.  In our simulations, we find a maximum $\rm Re = 7.6$. Low $\rm Re$ flows also indicate that the viscous forces are relevant. Viscous forces usually arise due to friction, for example, near the walls. The presence of both suspended and adhered droplets in a 3D channel leads to distinct profiles. In particular, the droplets adhered at the channel wall offer resistance to the flow of other droplets up to a certain point.

\begin{figure}[H]
\center
\includegraphics[width=1.0\linewidth]{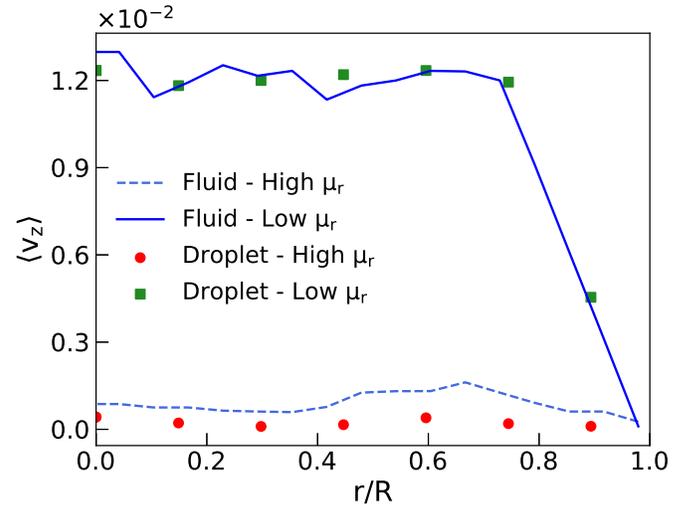}
\caption{Velocity profile for a system with $\gamma = 0.019$ before and after shear thinning. The forces are $ \rm F = 0.75 \times 10^{-6}$ and $ \rm F = 2.50 \times 10^{-6}$, respectively. The droplet and fluid velocity are plotted separately. Notice the peaks in the velocity of the surrounding fluid at high viscosity.}
\label{vprofile-fig}
\end{figure}

We plot the velocity distribution as a function of the nondimensional radius $r/R$. We measure the fluid and the droplet velocity separately before (high $\mu_r$) and after (low $\mu_r$) the discontinuous shear thinning. We take an average velocity across the channel length given by $\langle \mathbf{V_k} \rangle = \frac{\sum_{i=1}^{n} v_{ki}}{n}$ where $v_{ki}$ is the velocity at each node $i$ of the spatial discretization (of component $k$) and $n$ is the total number of nodes. The average is taken between  radial distance $r$ and $r+\Delta r$, for a fixed $\Delta r$. When shear thinning has occurred the velocity of the droplets and the surrounding fluid are similar and are consistent with plug flow behavior. We also observed that for a low number of droplets, no shear thinning occurs (plot not shown) as in 2D. The situation before shear thinning; however, contrasts with that of a 2D channel (with similar area fraction) where the system flows slowly and is in a nearly jammed state. While previous studies in 2D~\cite{Foglino2017,fei_2020} have not reported differences between fluid and droplet velocity before shear thinning, we observe differences in a 3D geometry. The velocity of the surrounding fluid is significantly higher than that of the droplets as shown in Fig.~\ref{vprofile-fig}. We point out that for a 3D channel with high volume fraction there is space between the droplets for the surrounding fluid to flow in contrast to the 2D case with a similar arrangement.
Due to the droplets velocity being very close to zero in our simulations, the droplets act as a soft porous medium~\cite{louvet_2010} with the surrounding fluid flowing past it. As we increase the forcing, the fluid eventually pushes the droplets leading to the discontinuous shear thinning. The droplets and fluid then flow with matching velocities. We highlight that, although simulations in 2D geometries can capture the essential flow qualities such as discontinuous shear thinning and plug flow behavior, 3D geometries give more realistic results. In particular, the difference between droplet and surrounding fluid velocities. The flow of the surrounding fluid between 2D droplets is affected by geometrical restrictions caused by narrow passages and dead-end pores between droplets~\cite{Koponen_1998,guariguata_2012}, especially at high area fractions. By contrast, 3D geometries with droplets, allows significant interconnected space between the droplets and therefore the surrounding fluid can flow through it. The importance of this effect and the role it plays in hydrodynamics of soft particles are not well understood.

\section{Summary and conclusion}
\label{conclusions-sec}

In conclusion,  we  have simulated the flow of soft particles in a 3D channel. For progressively higher values of the external force we observed discontinuous shear thinning. We studied in this transition, the effect of the surface tension, which is proportional to the particle deformability. At higher surface tension the droplets are less deformable and thus larger values of forcing are required for discontinuous shear thinning to occur. We observed that this transition occurs at a given capillary number. We also noticed that $\mu_r$ increases with the surface tension. This is to be expected as the particles become less deformable. Simulations for low volume fraction do not exhibit any discontinuous shear thinning in line with 2D studies. We analyzed the velocity profiles before and after shear thinning and noticed that after shear thinning the velocity profile is that of plug flow similar to 2D flows. However, closer inspection of the velocity profile before shear thinning revealed an essential feature that distinguishes our results from those in 2D. We noticed that the fluid and droplet velocities are different with the fluid velocity being higher and that of the droplets being near zero. This suggests that the droplets act as a soft porous medium. We propose that it would be useful to see how the viscosity ratio (between the droplets and the fluid) affects the overall flow.

To carry out this study we extended to 3D a previous multicomponent LBM that prevents coalescence between the droplets. We measured the disjoining pressure which is independent of the viscosity ratio. We were able to reduce spurious velocities which cause unphysical effects in the simulations such as sticking droplets and coalescence when collision occurs in certain directions. This was achieved by using higher order lattices in the streaming step. 

We note some prospects for experimental studies of deformable particles in microchannels. Recently, hydrodynamic resistance (the extra resistance due to the presence of an object in a channel) has been suggested as a parameter for characterizing such flows~\cite{sajeesh_2014}. It would be interesting to calculate the dependence of the hydrodynamic resistance on the capillary number in experiments with a similar setup to the one presented in this paper.

\section*{Conflicts of interest}
There are no conflicts to declare.

\section*{Acknowledgements}

We acknowledge financial support from the Portuguese Foundation for Science and Technology (FCT) under the contracts: PTDC/FIS-MAC/28146/2017 (LISBOA-01-0145-FEDER-028146), UIDB/00618/2020, UIDP/00618/2020 and 2020.08525.BD. 

\section*{Appendix:}
\begin{table}[h!]
\caption{Velocity vectors and weights for the D3Q41 lattice. The speed of sound $c_{s}^2$ is $1-\sqrt{2/5}$.}
\begin{tabular} {|c|c|}
\hline
 $\mathbf{c}_{i}$ & $w_{i}$ \\ \hline
 $(0,0,0)$ & $2(5045-1507 \sqrt{10}) / 2025$ \\
 $(\pm 1,0,0),(0,\pm 1,0),(0,0,\pm 1)$  & $377 /(5 \sqrt{10})-(91 / 40)$ \\
 $(\pm 1,\pm 1,0),(\pm 1,0,\pm 1),(0,\pm 1,\pm 1)$ & $(55-17 \sqrt{10}) / 50$ \\
 $(\pm 1,\pm 1,\pm 1)$ & (233 $\sqrt{10}-730) / 1600$ \\
 $(\pm 3,0,0),(0,\pm 3,0),(0,0,\pm 3)$  & $(295-92 \sqrt{10}) / 16200$ \\ 
 $(\pm 3,\pm 3,\pm 3)$  & $(130-41 \sqrt{10}) / 129600$ \\ \hline
\end{tabular}
\label{d3q41-table}
\end{table}

\begin{table}[h!]
\caption{Velocity vectors and weights for the D3Q39 lattice. The speed of sound $c_{s}^2$ is $2/3$.}
\begin{tabular} {|c|c|}
\hline
$\mathbf{c}_{i}$ & $w_{i}$\\ \hline
 $(0,0,0)$ & 1/12 \\ 
 $(\pm 1,0,0)$  & 1/12 \\
 $(\pm 1,\pm 1,\pm 1)$ & 1/27 \\
 $(\pm 2,0,0)$  & 2/135 \\
 $(\pm 2,\pm 2,0)$ & 1/142 \\
 $(\pm 3,0,0)$  & 1/1620 \\ \hline
\end{tabular}
\label{d3q39-table}
\end{table}





\bibliography{rsc} 

\providecommand*{\mcitethebibliography}{\thebibliography}
\csname @ifundefined\endcsname{endmcitethebibliography}
{\let\endmcitethebibliography\endthebibliography}{}
\begin{mcitethebibliography}{52}
\providecommand*{\natexlab}[1]{#1}
\providecommand*{\mciteSetBstSublistMode}[1]{}
\providecommand*{\mciteSetBstMaxWidthForm}[2]{}
\providecommand*{\mciteBstWouldAddEndPuncttrue}
  {\def\EndOfBibitem{\unskip.}}
\providecommand*{\mciteBstWouldAddEndPunctfalse}
  {\let\EndOfBibitem\relax}
\providecommand*{\mciteSetBstMidEndSepPunct}[3]{}
\providecommand*{\mciteSetBstSublistLabelBeginEnd}[3]{}
\providecommand*{\EndOfBibitem}{}
\mciteSetBstSublistMode{f}
\mciteSetBstMaxWidthForm{subitem}
{(\emph{\alph{mcitesubitemcount}})}
\mciteSetBstSublistLabelBeginEnd{\mcitemaxwidthsubitemform\space}
{\relax}{\relax}

\bibitem[Xu \emph{et~al.}(2009)Xu, Hashimoto, Dang, Hoare, Kohane, Whitesides,
  Langer, and Anderson]{Xu_2009}
Q.~Xu, M.~Hashimoto, T.~T. Dang, T.~Hoare, D.~S. Kohane, G.~M. Whitesides,
  R.~Langer and D.~G. Anderson, \emph{Small}, 2009, \textbf{5},
  1575--1581\relax
\mciteBstWouldAddEndPuncttrue
\mciteSetBstMidEndSepPunct{\mcitedefaultmidpunct}
{\mcitedefaultendpunct}{\mcitedefaultseppunct}\relax
\EndOfBibitem
\bibitem[Li \emph{et~al.}(2014)Li, Chen, Liu, Lu, and Fu]{li_2014}
X.~Li, W.~Chen, G.~Liu, W.~Lu and J.~Fu, \emph{Lab Chip}, 2014, \textbf{14},
  2565--2575\relax
\mciteBstWouldAddEndPuncttrue
\mciteSetBstMidEndSepPunct{\mcitedefaultmidpunct}
{\mcitedefaultendpunct}{\mcitedefaultseppunct}\relax
\EndOfBibitem
\bibitem[Young and Beebe(2010)]{young_fundamentals_2010}
E.~W.~K. Young and D.~J. Beebe, \emph{Chemical Society Reviews}, 2010,
  \textbf{39}, 1036--1048\relax
\mciteBstWouldAddEndPuncttrue
\mciteSetBstMidEndSepPunct{\mcitedefaultmidpunct}
{\mcitedefaultendpunct}{\mcitedefaultseppunct}\relax
\EndOfBibitem
\bibitem[Mehling and Tay(2014)]{mehling_microfluidic_2014}
M.~Mehling and S.~Tay, \emph{Current Opinion in Biotechnology}, 2014,
  \textbf{25}, 95--102\relax
\mciteBstWouldAddEndPuncttrue
\mciteSetBstMidEndSepPunct{\mcitedefaultmidpunct}
{\mcitedefaultendpunct}{\mcitedefaultseppunct}\relax
\EndOfBibitem
\bibitem[Koumakis \emph{et~al.}(2012)Koumakis, Pamvouxoglou, Poulos, and
  Petekidis]{koumakis2012}
N.~Koumakis, A.~Pamvouxoglou, A.~S. Poulos and G.~Petekidis, \emph{Soft
  Matter}, 2012, \textbf{8}, 4271--4284\relax
\mciteBstWouldAddEndPuncttrue
\mciteSetBstMidEndSepPunct{\mcitedefaultmidpunct}
{\mcitedefaultendpunct}{\mcitedefaultseppunct}\relax
\EndOfBibitem
\bibitem[Vaart \emph{et~al.}(2013)Vaart, Rahmani, Zargar, Bonn, and
  Schall]{vaart2013}
K.~v.~d. Vaart, Y.~Rahmani, R.~Zargar, D.~Bonn and P.~Schall, \emph{Journal of
  Rheology}, 2013, \textbf{57}, 1195\relax
\mciteBstWouldAddEndPuncttrue
\mciteSetBstMidEndSepPunct{\mcitedefaultmidpunct}
{\mcitedefaultendpunct}{\mcitedefaultseppunct}\relax
\EndOfBibitem
\bibitem[Hou \emph{et~al.}(2001)Hou, Lowengrub, and Shelley]{Hou_2001}
T.~Hou, J.~Lowengrub and M.~Shelley, \emph{Journal of Computational Physics},
  2001, \textbf{169}, 302--362\relax
\mciteBstWouldAddEndPuncttrue
\mciteSetBstMidEndSepPunct{\mcitedefaultmidpunct}
{\mcitedefaultendpunct}{\mcitedefaultseppunct}\relax
\EndOfBibitem
\bibitem[Pozrikidis(2001)]{POZRIKIDIS_2001}
C.~Pozrikidis, \emph{Journal of Computational Physics}, 2001, \textbf{169}, 250
  -- 301\relax
\mciteBstWouldAddEndPuncttrue
\mciteSetBstMidEndSepPunct{\mcitedefaultmidpunct}
{\mcitedefaultendpunct}{\mcitedefaultseppunct}\relax
\EndOfBibitem
\bibitem[Zhou and Pozrikidis(1994)]{Zhou1994}
H.~Zhou and C.~Pozrikidis, \emph{Physics of Fluids}, 1994, \textbf{6},
  80--94\relax
\mciteBstWouldAddEndPuncttrue
\mciteSetBstMidEndSepPunct{\mcitedefaultmidpunct}
{\mcitedefaultendpunct}{\mcitedefaultseppunct}\relax
\EndOfBibitem
\bibitem[Nott and Brady(1994)]{nott_brady_1994}
P.~R. Nott and J.~F. Brady, \emph{Journal of Fluid Mechanics}, 1994,
  \textbf{275}, 157–199\relax
\mciteBstWouldAddEndPuncttrue
\mciteSetBstMidEndSepPunct{\mcitedefaultmidpunct}
{\mcitedefaultendpunct}{\mcitedefaultseppunct}\relax
\EndOfBibitem
\bibitem[D{\"u}nweg and Ladd(2009)]{Dunweg_2009}
B.~D{\"u}nweg and A.~J.~C. Ladd, in \emph{Lattice Boltzmann Simulations of Soft
  Matter Systems}, ed. C.~Holm and K.~Kremer, Springer Berlin Heidelberg,
  Berlin, Heidelberg, 2009, pp. 89--166\relax
\mciteBstWouldAddEndPuncttrue
\mciteSetBstMidEndSepPunct{\mcitedefaultmidpunct}
{\mcitedefaultendpunct}{\mcitedefaultseppunct}\relax
\EndOfBibitem
\bibitem[Krüger \emph{et~al.}(2011)Krüger, Varnik, and Raabe]{kruger_2011}
T.~Krüger, F.~Varnik and D.~Raabe, \emph{Computers \& Mathematics with
  Applications}, 2011, \textbf{61}, 3485 -- 3505\relax
\mciteBstWouldAddEndPuncttrue
\mciteSetBstMidEndSepPunct{\mcitedefaultmidpunct}
{\mcitedefaultendpunct}{\mcitedefaultseppunct}\relax
\EndOfBibitem
\bibitem[Chen(2014)]{chen_2014}
Y.-L. Chen, \emph{RSC Adv.}, 2014, \textbf{4}, 17908--17916\relax
\mciteBstWouldAddEndPuncttrue
\mciteSetBstMidEndSepPunct{\mcitedefaultmidpunct}
{\mcitedefaultendpunct}{\mcitedefaultseppunct}\relax
\EndOfBibitem
\bibitem[Krüger \emph{et~al.}(2014)Krüger, Kaoui, and
  Harting]{kruger_kaoui_harting_2014}
T.~Krüger, B.~Kaoui and J.~Harting, \emph{Journal of Fluid Mechanics}, 2014,
  \textbf{751}, 725–745\relax
\mciteBstWouldAddEndPuncttrue
\mciteSetBstMidEndSepPunct{\mcitedefaultmidpunct}
{\mcitedefaultendpunct}{\mcitedefaultseppunct}\relax
\EndOfBibitem
\bibitem[Brown and Jaeger(2009)]{brown_2009}
E.~Brown and H.~M. Jaeger, \emph{Phys. Rev. Lett.}, 2009, \textbf{103},
  086001\relax
\mciteBstWouldAddEndPuncttrue
\mciteSetBstMidEndSepPunct{\mcitedefaultmidpunct}
{\mcitedefaultendpunct}{\mcitedefaultseppunct}\relax
\EndOfBibitem
\bibitem[Fall \emph{et~al.}(2015)Fall, Bertrand, Hautemayou, Mezi\`ere,
  Moucheront, Lema\^{\i}tre, and Ovarlez]{fall_2015}
A.~Fall, F.~Bertrand, D.~Hautemayou, C.~Mezi\`ere, P.~Moucheront,
  A.~Lema\^{\i}tre and G.~Ovarlez, \emph{Phys. Rev. Lett.}, 2015, \textbf{114},
  098301\relax
\mciteBstWouldAddEndPuncttrue
\mciteSetBstMidEndSepPunct{\mcitedefaultmidpunct}
{\mcitedefaultendpunct}{\mcitedefaultseppunct}\relax
\EndOfBibitem
\bibitem[Zarraga \emph{et~al.}(2000)Zarraga, Hill, and Leighton]{isidro}
I.~E. Zarraga, D.~A. Hill and D.~T. Leighton, \emph{Journal of Rheology}, 2000,
  \textbf{44}, 185--220\relax
\mciteBstWouldAddEndPuncttrue
\mciteSetBstMidEndSepPunct{\mcitedefaultmidpunct}
{\mcitedefaultendpunct}{\mcitedefaultseppunct}\relax
\EndOfBibitem
\bibitem[Dai \emph{et~al.}(2013)Dai, Bertevas, Qi, and Tanner]{shao_2013}
S.-C. Dai, E.~Bertevas, F.~Qi and R.~I. Tanner, \emph{Journal of Rheology},
  2013, \textbf{57}, 493--510\relax
\mciteBstWouldAddEndPuncttrue
\mciteSetBstMidEndSepPunct{\mcitedefaultmidpunct}
{\mcitedefaultendpunct}{\mcitedefaultseppunct}\relax
\EndOfBibitem
\bibitem[Lin \emph{et~al.}(2014)Lin, Tan, Phan-Thien, and Khoo]{lin_2014}
Y.~Lin, G.~W.~H. Tan, N.~Phan-Thien and B.~C. Khoo, \emph{Journal of
  Non-Newtonian Fluid Mechanics}, 2014, \textbf{212}, 13 -- 17\relax
\mciteBstWouldAddEndPuncttrue
\mciteSetBstMidEndSepPunct{\mcitedefaultmidpunct}
{\mcitedefaultendpunct}{\mcitedefaultseppunct}\relax
\EndOfBibitem
\bibitem[SINGH and NOTT(2003)]{singh_nott_2003}
A.~SINGH and P.~R. NOTT, \emph{Journal of Fluid Mechanics}, 2003, \textbf{490},
  293–320\relax
\mciteBstWouldAddEndPuncttrue
\mciteSetBstMidEndSepPunct{\mcitedefaultmidpunct}
{\mcitedefaultendpunct}{\mcitedefaultseppunct}\relax
\EndOfBibitem
\bibitem[Otsubo and Prud'homme(1994)]{otsubo_effect_1994}
Y.~Otsubo and R.~K. Prud'homme, \emph{Rheologica Acta}, 1994, \textbf{33},
  303--306\relax
\mciteBstWouldAddEndPuncttrue
\mciteSetBstMidEndSepPunct{\mcitedefaultmidpunct}
{\mcitedefaultendpunct}{\mcitedefaultseppunct}\relax
\EndOfBibitem
\bibitem[Pal(1996)]{Pal1996}
R.~Pal, \emph{AIChE Journal}, 1996, \textbf{42}, 3181--3190\relax
\mciteBstWouldAddEndPuncttrue
\mciteSetBstMidEndSepPunct{\mcitedefaultmidpunct}
{\mcitedefaultendpunct}{\mcitedefaultseppunct}\relax
\EndOfBibitem
\bibitem[Pal(2000)]{Pal2000}
R.~Pal, \emph{Journal of Colloid and Interface Science}, 2000, \textbf{225},
  359 -- 366\relax
\mciteBstWouldAddEndPuncttrue
\mciteSetBstMidEndSepPunct{\mcitedefaultmidpunct}
{\mcitedefaultendpunct}{\mcitedefaultseppunct}\relax
\EndOfBibitem
\bibitem[Loewenberg and Hinch(1996)]{loewenberg_hinch_1996}
M.~Loewenberg and E.~J. Hinch, \emph{Journal of Fluid Mechanics}, 1996,
  \textbf{321}, 395–419\relax
\mciteBstWouldAddEndPuncttrue
\mciteSetBstMidEndSepPunct{\mcitedefaultmidpunct}
{\mcitedefaultendpunct}{\mcitedefaultseppunct}\relax
\EndOfBibitem
\bibitem[Doddi and Bagchi(2009)]{Doddi2009}
S.~K. Doddi and P.~Bagchi, \emph{Phys. Rev. E}, 2009, \textbf{79}, 046318\relax
\mciteBstWouldAddEndPuncttrue
\mciteSetBstMidEndSepPunct{\mcitedefaultmidpunct}
{\mcitedefaultendpunct}{\mcitedefaultseppunct}\relax
\EndOfBibitem
\bibitem[Lázaro \emph{et~al.}(2014)Lázaro, Hernández-Machado, and
  Pagonabarraga]{lazaro2014_I}
G.~R. Lázaro, A.~Hernández-Machado and I.~Pagonabarraga, \emph{Soft Matter},
  2014, \textbf{10}, 7195--7206\relax
\mciteBstWouldAddEndPuncttrue
\mciteSetBstMidEndSepPunct{\mcitedefaultmidpunct}
{\mcitedefaultendpunct}{\mcitedefaultseppunct}\relax
\EndOfBibitem
\bibitem[Lázaro \emph{et~al.}(2014)Lázaro, Hernández-Machado, and
  Pagonabarraga]{lazaro2014_II}
G.~R. Lázaro, A.~Hernández-Machado and I.~Pagonabarraga, \emph{Soft Matter},
  2014, \textbf{10}, 7207--7217\relax
\mciteBstWouldAddEndPuncttrue
\mciteSetBstMidEndSepPunct{\mcitedefaultmidpunct}
{\mcitedefaultendpunct}{\mcitedefaultseppunct}\relax
\EndOfBibitem
\bibitem[Foglino \emph{et~al.}(2017)Foglino, Morozov, Henrich, and
  Marenduzzo]{Foglino2017}
M.~Foglino, A.~N. Morozov, O.~Henrich and D.~Marenduzzo, \emph{Phys. Rev.
  Lett.}, 2017, \textbf{119}, 208002\relax
\mciteBstWouldAddEndPuncttrue
\mciteSetBstMidEndSepPunct{\mcitedefaultmidpunct}
{\mcitedefaultendpunct}{\mcitedefaultseppunct}\relax
\EndOfBibitem
\bibitem[Fei \emph{et~al.}(2020)Fei, Scagliarini, Luo, and Succi]{fei_2020}
L.~Fei, A.~Scagliarini, K.~H. Luo and S.~Succi, \emph{Soft Matter}, 2020,
  \textbf{16}, 651--658\relax
\mciteBstWouldAddEndPuncttrue
\mciteSetBstMidEndSepPunct{\mcitedefaultmidpunct}
{\mcitedefaultendpunct}{\mcitedefaultseppunct}\relax
\EndOfBibitem
\bibitem[Fedosov \emph{et~al.}(2014)Fedosov, Peltomäki, and
  Gompper]{fedosov_2014}
D.~A. Fedosov, M.~Peltomäki and G.~Gompper, \emph{Soft Matter}, 2014,
  \textbf{10}, 4258--4267\relax
\mciteBstWouldAddEndPuncttrue
\mciteSetBstMidEndSepPunct{\mcitedefaultmidpunct}
{\mcitedefaultendpunct}{\mcitedefaultseppunct}\relax
\EndOfBibitem
\bibitem[Benzi \emph{et~al.}(2009)Benzi, Chibbaro, and
  Succi]{benzi_mesoscopic_2009}
R.~Benzi, S.~Chibbaro and S.~Succi, \emph{Physical Review Letters}, 2009,
  \textbf{102}, 026002\relax
\mciteBstWouldAddEndPuncttrue
\mciteSetBstMidEndSepPunct{\mcitedefaultmidpunct}
{\mcitedefaultendpunct}{\mcitedefaultseppunct}\relax
\EndOfBibitem
\bibitem[Guo \emph{et~al.}(2002)Guo, Zheng, and Shi]{Guo_2002}
Z.~Guo, C.~Zheng and B.~Shi, \emph{Phys. Rev. E}, 2002, \textbf{65},
  046308\relax
\mciteBstWouldAddEndPuncttrue
\mciteSetBstMidEndSepPunct{\mcitedefaultmidpunct}
{\mcitedefaultendpunct}{\mcitedefaultseppunct}\relax
\EndOfBibitem
\bibitem[Guo \emph{et~al.}(2002)Guo, Zheng, and Shi]{guo_zheng_shi_2002}
Z.~Guo, C.~Zheng and B.~Shi, \emph{Physical Review E}, 2002, \textbf{65},
  046308\relax
\mciteBstWouldAddEndPuncttrue
\mciteSetBstMidEndSepPunct{\mcitedefaultmidpunct}
{\mcitedefaultendpunct}{\mcitedefaultseppunct}\relax
\EndOfBibitem
\bibitem[Krueger \emph{et~al.}(2016)Krueger, Kusumaatmaja, Kuzmin, Shardt,
  Silva, and Viggen]{kruger2016}
T.~Krueger, H.~Kusumaatmaja, A.~Kuzmin, O.~Shardt, G.~Silva and E.~Viggen,
  \emph{The Lattice Boltzmann Method: Principles and Practice}, Springer,
  2016\relax
\mciteBstWouldAddEndPuncttrue
\mciteSetBstMidEndSepPunct{\mcitedefaultmidpunct}
{\mcitedefaultendpunct}{\mcitedefaultseppunct}\relax
\EndOfBibitem
\bibitem[Falcucci \emph{et~al.}(2007)Falcucci, Bella, Shiatti, Chibbaro,
  Sbragaglia, and Succi]{falcucci_lattice_2007}
G.~Falcucci, G.~Bella, G.~Shiatti, S.~Chibbaro, M.~Sbragaglia and S.~Succi,
  \emph{Communications in Computational Physics}, 2007, \textbf{2},
  1071--1084\relax
\mciteBstWouldAddEndPuncttrue
\mciteSetBstMidEndSepPunct{\mcitedefaultmidpunct}
{\mcitedefaultendpunct}{\mcitedefaultseppunct}\relax
\EndOfBibitem
\bibitem[Dollet \emph{et~al.}(2015)Dollet, Scagliarini, and
  Sbragaglia]{dollet_two-dimensional_2015}
B.~Dollet, A.~Scagliarini and M.~Sbragaglia, \emph{Journal of Fluid Mechanics},
  2015, \textbf{766}, 556--589\relax
\mciteBstWouldAddEndPuncttrue
\mciteSetBstMidEndSepPunct{\mcitedefaultmidpunct}
{\mcitedefaultendpunct}{\mcitedefaultseppunct}\relax
\EndOfBibitem
\bibitem[Falcucci \emph{et~al.}(2007)Falcucci, Bella, Shiatti, Chibbaro,
  Sbragaglia, and Succi]{falcucci_2007}
G.~Falcucci, G.~Bella, G.~Shiatti, S.~Chibbaro, M.~Sbragaglia and S.~Succi,
  \emph{Communications in Computational Physics - COMMUN COMPUT PHYS}, 2007,
  \textbf{2}, 1071--1084\relax
\mciteBstWouldAddEndPuncttrue
\mciteSetBstMidEndSepPunct{\mcitedefaultmidpunct}
{\mcitedefaultendpunct}{\mcitedefaultseppunct}\relax
\EndOfBibitem
\bibitem[Benzi \emph{et~al.}(2009)Benzi, Sbragaglia, Succi, Bernaschi, and
  Chibbaro]{benzi_2007}
R.~Benzi, M.~Sbragaglia, S.~Succi, M.~Bernaschi and S.~Chibbaro, \emph{The
  Journal of Chemical Physics}, 2009, \textbf{131}, 104903\relax
\mciteBstWouldAddEndPuncttrue
\mciteSetBstMidEndSepPunct{\mcitedefaultmidpunct}
{\mcitedefaultendpunct}{\mcitedefaultseppunct}\relax
\EndOfBibitem
\bibitem[Sbragaglia \emph{et~al.}(2007)Sbragaglia, Benzi, Biferale, Succi,
  Sugiyama, and Toschi]{sbragalia_2007}
M.~Sbragaglia, R.~Benzi, L.~Biferale, S.~Succi, K.~Sugiyama and F.~Toschi,
  \emph{Phys. Rev. E}, 2007, \textbf{75}, 026702\relax
\mciteBstWouldAddEndPuncttrue
\mciteSetBstMidEndSepPunct{\mcitedefaultmidpunct}
{\mcitedefaultendpunct}{\mcitedefaultseppunct}\relax
\EndOfBibitem
\bibitem[Peng \emph{et~al.}(2019)Peng, Ayala, Ayala, and
  Wang]{peng_isotropy_2019}
C.~Peng, L.~F. Ayala, O.~M. Ayala and L.-P. Wang, \emph{Computers \& Fluids},
  2019, \textbf{191}, 104257\relax
\mciteBstWouldAddEndPuncttrue
\mciteSetBstMidEndSepPunct{\mcitedefaultmidpunct}
{\mcitedefaultendpunct}{\mcitedefaultseppunct}\relax
\EndOfBibitem
\bibitem[Krüger \emph{et~al.}(2017)Krüger, Kusumaatmaja, Kuzmin, Shardt,
  Silva, and Viggen]{kruger_lattice_2017}
T.~Krüger, H.~Kusumaatmaja, A.~Kuzmin, O.~Shardt, G.~Silva and E.~M. Viggen,
  \emph{The {Lattice} {Boltzmann} {Method}}, Springer International Publishing,
  Cham, 2017\relax
\mciteBstWouldAddEndPuncttrue
\mciteSetBstMidEndSepPunct{\mcitedefaultmidpunct}
{\mcitedefaultendpunct}{\mcitedefaultseppunct}\relax
\EndOfBibitem
\bibitem[Sbragaglia \emph{et~al.}(2012)Sbragaglia, Benzi, Bernaschi, and
  Succi]{sbragaglia_emergence_2012}
M.~Sbragaglia, R.~Benzi, M.~Bernaschi and S.~Succi, \emph{Soft Matter}, 2012,
  \textbf{8}, 10773\relax
\mciteBstWouldAddEndPuncttrue
\mciteSetBstMidEndSepPunct{\mcitedefaultmidpunct}
{\mcitedefaultendpunct}{\mcitedefaultseppunct}\relax
\EndOfBibitem
\bibitem[Benzi \emph{et~al.}(2015)Benzi, Sbragaglia, Scagliarini, Perlekar,
  Bernaschi, Succi, and Toschi]{benzi_internal_2015}
R.~Benzi, M.~Sbragaglia, A.~Scagliarini, P.~Perlekar, M.~Bernaschi, S.~Succi
  and F.~Toschi, \emph{Soft Matter}, 2015, \textbf{11}, 1271--1280\relax
\mciteBstWouldAddEndPuncttrue
\mciteSetBstMidEndSepPunct{\mcitedefaultmidpunct}
{\mcitedefaultendpunct}{\mcitedefaultseppunct}\relax
\EndOfBibitem
\bibitem[Fei \emph{et~al.}(2018)Fei, Scagliarini, Montessori, Lauricella,
  Succi, and Luo]{fei_mesoscopic_2018}
L.~Fei, A.~Scagliarini, A.~Montessori, M.~Lauricella, S.~Succi and K.~H. Luo,
  \emph{Physical Review Fluids}, 2018, \textbf{3}, 104304\relax
\mciteBstWouldAddEndPuncttrue
\mciteSetBstMidEndSepPunct{\mcitedefaultmidpunct}
{\mcitedefaultendpunct}{\mcitedefaultseppunct}\relax
\EndOfBibitem
\bibitem[Lanotte \emph{et~al.}(2016)Lanotte, Mauer, Mendez, Fedosov, Fromental,
  Claveria, Nicoud, Gompper, and Abkarian]{lanotte_red_2016}
L.~Lanotte, J.~Mauer, S.~Mendez, D.~A. Fedosov, J.-M. Fromental, V.~Claveria,
  F.~Nicoud, G.~Gompper and M.~Abkarian, \emph{Proceedings of the National
  Academy of Sciences}, 2016, \textbf{113}, 13289--13294\relax
\mciteBstWouldAddEndPuncttrue
\mciteSetBstMidEndSepPunct{\mcitedefaultmidpunct}
{\mcitedefaultendpunct}{\mcitedefaultseppunct}\relax
\EndOfBibitem
\bibitem[Kroupa \emph{et~al.}(2017)Kroupa, Soos, and Kosek]{kroupa_slip_2017}
M.~Kroupa, M.~Soos and J.~Kosek, \emph{Phys. Chem. Chem. Phys.}, 2017,
  \textbf{19}, 5979--5984\relax
\mciteBstWouldAddEndPuncttrue
\mciteSetBstMidEndSepPunct{\mcitedefaultmidpunct}
{\mcitedefaultendpunct}{\mcitedefaultseppunct}\relax
\EndOfBibitem
\bibitem[Cwalina and Wagner(2014)]{Cwalina_2014}
C.~D. Cwalina and N.~J. Wagner, \emph{Journal of Rheology}, 2014, \textbf{58},
  949--967\relax
\mciteBstWouldAddEndPuncttrue
\mciteSetBstMidEndSepPunct{\mcitedefaultmidpunct}
{\mcitedefaultendpunct}{\mcitedefaultseppunct}\relax
\EndOfBibitem
\bibitem[Schlichting~(Deceased) and Gersten(2017)]{Schlichting(Deceased)2017}
H.~Schlichting~(Deceased) and K.~Gersten, in \emph{Exact Solutions of the
  Navier--Stokes Equations}, Springer Berlin Heidelberg, Berlin, Heidelberg,
  2017, pp. 101--142\relax
\mciteBstWouldAddEndPuncttrue
\mciteSetBstMidEndSepPunct{\mcitedefaultmidpunct}
{\mcitedefaultendpunct}{\mcitedefaultseppunct}\relax
\EndOfBibitem
\bibitem[Louvet \emph{et~al.}(2010)Louvet, H\"ohler, and Pitois]{louvet_2010}
N.~Louvet, R.~H\"ohler and O.~Pitois, \emph{Phys. Rev. E}, 2010, \textbf{82},
  041405\relax
\mciteBstWouldAddEndPuncttrue
\mciteSetBstMidEndSepPunct{\mcitedefaultmidpunct}
{\mcitedefaultendpunct}{\mcitedefaultseppunct}\relax
\EndOfBibitem
\bibitem[Koponen \emph{et~al.}(1997)Koponen, Kataja, and Timonen]{Koponen_1998}
A.~Koponen, M.~Kataja and J.~Timonen, \emph{Phys. Rev. E}, 1997, \textbf{56},
  3319--3325\relax
\mciteBstWouldAddEndPuncttrue
\mciteSetBstMidEndSepPunct{\mcitedefaultmidpunct}
{\mcitedefaultendpunct}{\mcitedefaultseppunct}\relax
\EndOfBibitem
\bibitem[Guariguata \emph{et~al.}(2012)Guariguata, Pascall, Gilmer, Sum, Sloan,
  Koh, and Wu]{guariguata_2012}
A.~Guariguata, M.~A. Pascall, M.~W. Gilmer, A.~K. Sum, E.~D. Sloan, C.~A. Koh
  and D.~T. Wu, \emph{Phys. Rev. E}, 2012, \textbf{86}, 061311\relax
\mciteBstWouldAddEndPuncttrue
\mciteSetBstMidEndSepPunct{\mcitedefaultmidpunct}
{\mcitedefaultendpunct}{\mcitedefaultseppunct}\relax
\EndOfBibitem
\bibitem[Sajeesh \emph{et~al.}(2014)Sajeesh, Doble, and Sen]{sajeesh_2014}
P.~Sajeesh, M.~Doble and A.~K. Sen, \emph{Biomicrofluidics}, 2014, \textbf{8},
  054112\relax
\mciteBstWouldAddEndPuncttrue
\mciteSetBstMidEndSepPunct{\mcitedefaultmidpunct}
{\mcitedefaultendpunct}{\mcitedefaultseppunct}\relax
\EndOfBibitem
\end{mcitethebibliography}
\bibliographystyle{rsc} 
\end{document}